\documentclass[pra,twocolumn,preprintnumbers,amsmath,amssymb,nofootinbib,floatfix]{revtex4}

\usepackage{graphicx, bm, tikz}
\usepackage[breaklinks]{hyperref}

\makeatletter
\def\graphicscale{\twocolumn@sw{0.3}{0.4}}
\def\graphicthreescale{\twocolumn@sw{0.3}{0.4}}

\begin{document}

\title{Competing coherent and dissipative dynamics close to quantum criticality}

\author{Davide Nigro}
\affiliation{Dipartimento di Fisica dell'Universit\`a di Pisa
        and INFN, Largo Pontecorvo 3, I-56127 Pisa, Italy}

\author{Davide Rossini}
\affiliation{Dipartimento di Fisica dell'Universit\`a di Pisa
        and INFN, Largo Pontecorvo 3, I-56127 Pisa, Italy}

\author{Ettore Vicari} 
\altaffiliation{Authors are listed in alphabetic order.}
\affiliation{Dipartimento di Fisica dell'Universit\`a di Pisa
        and INFN, Largo Pontecorvo 3, I-56127 Pisa, Italy}

\date{\today}

\begin{abstract}
  We investigate the competition of coherent and dissipative dynamics
  in many-body systems at continuous quantum transitions.  We consider
  dissipative mechanisms that can be effectively described by Lindblad
  equations for the density matrix of the system.  The interplay
  between the critical coherent dynamics and dissipation is addressed
  within a dynamic finite-size scaling framework, which allows us to
  identify the regime where they develop a nontrivial competition.  We
  analyze protocols that start from critical many-body ground states,
  and put forward general dynamic scaling behaviors involving
  the Hamiltonian parameters and the coupling associated with the
  dissipation.  This scaling scenario is supported by a numerical
  study of the dynamic behavior of a one-dimensional lattice fermion
  gas undergoing a quantum Ising transition, in the presence of
  dissipative mechanisms such as local pumping, decaying and
  dephasing.
\end{abstract}

\maketitle

% ========================= BODY =========================

\section{Introduction}
\label{sec:Introduction}

Understanding the quantum dynamics of many-body systems is one of the
greatest challenges of modern physics. The recent progress in atomic
physics and quantum optical technologies has provided a great
opportunity for a thorough investigation of the interplay between the
coherent quantum dynamics and the (practically unavoidable)
dissipative effects, due to the interaction with the
environment~\cite{HTK-12, MDPZ-12, CC-13, AKM-14}.  Likely, the most
intricate regime is the one characterized by an actual competition of
both dynamic mechanisms, which may develop a nontrivial interplay.
This can be responsible for the emergence of further interesting
phenomena in many-body systems, in particular when they are close to a
quantum phase transition, where quantum critical fluctuations emerge
and correlations develop a diverging length scale~\cite{Sachdev-book}.

In this paper we study the dynamics of open {\em critical} many-body
systems, whose Hamiltonians are close to a continuous quantum critical
point.  We consider a class of dissipative mechanisms that can be
effectively described by Lindblad equations for the density matrix of
the system~\cite{BP-book, RH-book}.  We address the interplay between
the critical coherent dynamics and dissipative mechanisms, by
considering dynamic protocols that start from ground states, or
low-temperature Gibbs distributions, close to quantum transitions.
Our approach exploits a dynamic finite-size scaling (FSS) framework,
which accounts for both the critical Hamiltonian and dissipation
drivings, and allows us to identify the dynamic regime where a
nontrivial competition develops.  General scaling behaviors are put
forward, involving both the Hamiltonian parameters and the couplings
associated with the dissipative terms.  We thus achieve the notable
result of combining intrinsically
different dynamic mechanisms in a unique framework.

To verify the emerging scaling scenario, we consider the paradigmatic
one-dimensional Kitaev fermion model~\cite{Kitaev-01}.  We study its
dynamic behavior close to its quantum Ising transition, in the
presence of local incoherent pumping, decay and dephasing. 
Numerical results reported below nicely support the general dynamic
FSS theory addressing the competition between critical coherent
dynamics and dissipation.

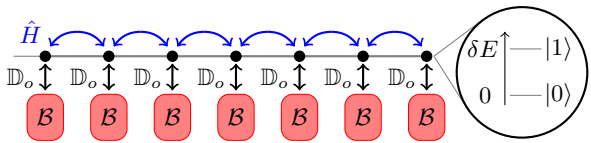
\begin{figure}[!h]
  \begin{tikzpicture}[scale=0.65]
     \draw [gray,thin] (0.65,0)--(9.1,0);
     \draw [gray,thick] (1.3,0)--(9.1,0);

% Energy levels
	\draw [<-] (10.7,0.5)--(10.7,-1);
	 
	 \draw [gray] (10.8,-0.8) --(11.45,-0.8);
	 \node at (10.3,-0.8) {0};
	 \node at (11.8,-0.8) {$\vert 0\rangle$};
	 
	 \draw [gray] (10.8,0.15) --(11.45,0.15);
	 \node at (10.25,0.15) {$\delta E$};
	 \node at (11.8,0.15) {$\vert 1\rangle$};

     \draw [thin,gray] (9.25,0)--(10,-1.18);
     \draw [thin,gray] (9.25,0)--(10.25,0.75);
     \draw[thick] (11.05,-0.33) circle (38 pt);    

     \node at (1,0.5) {\textcolor{blue}{$\hat{H}$}};
     
 \foreach \i in {1,...,6}
{
	\draw [<-,blue,thick] ({\i*1.3+0.1},0.2) to [out=55,in=180] ({\i*1.3+0.65},0.5) ;
	\draw [->,blue,thick] ({\i*1.3+0.65},0.5) to [out=0,in=125] ({\i*1.3+1.2},0.2) ;
}
    \foreach \i in {1,...,7}
{
        \draw[thick,<->,black] ({\i*1.3},-0.15)--({\i*1.3},-0.7);
        \filldraw [black] ({\i*1.3},0) circle (3pt);
        \draw [very thick, red, rounded corners]({\i*1.3-0.35},-1.7) rectangle ({\i*1.3+0.35},-0.8);
        \filldraw[red!50!white, rounded corners] ({\i*1.3-0.35},-1.7) rectangle ({\i*1.3+0.35},-0.8);
        \node at ({\i*1.3},-1.25) {$\mathcal{B}$};
        \node at ({\i*1.3-0.5},-0.45) { \textcolor{black}{$\mathbb{D}_o $}};
}
  \end{tikzpicture}
  \caption{Sketch of a Fermi lattice gas in one dimension.
    Particles undergo coherent pairing and tunneling mechanisms (bidirectional
    blue arrows) between neighboring lattice sites (black dots).
    The bubble indicates the two-level nature of each site,
    with $\delta E$ denoting the corresponding onsite energy spacing,
    [$\delta E = |\mu|$, for model~\eqref{kitaev2}].
    Each site is homogeneously and weakly coupled to an external
    and independent bath ${\mathcal B}$ (vertical black arrows),
    whose effect is to introduce local incoherent particle
    losses, pumping, or dephasing.}
  \label{fig:sketch}
\end{figure}

Our considerations apply to a generic $d$-dimensional many-body system
with Hamiltonian $\hat H$, close to a zero-temperature transition
driven by quantum fluctuations~\cite{Sachdev-book, SGCS-97}.  A
quantum transition is generally characterized by few relevant
perturbations, whose tuning gives rise to quantum critical behaviors,
characterized by a diverging length scale and universal power laws.
However, these features generally disappear in the presence of dissipation.
We assume that the many-body system also interacts with the environment,
so that the time dependence of its
density matrix $\rho$ is described by the Lindblad master
equation~\cite{BP-book}
\begin{equation}
  {\partial\rho\over \partial t} = -{i\over \hslash}[ \hat H,\rho]
  + u \sum_o {\mathbb D}_o[\rho]\,,
  \label{lindblaseq}
\end{equation}
where the first term provides the coherent driving, while the second
term accounts for the coupling to the environment.  Its form depends
on the nature of the dissipation arising from the interaction with the
bath, which is effectively described by a set of dissipators
${\mathbb D}_o$, and a global coupling $u>0$.  In the case of weak
coupling to Markovian baths, the trace-preserving superoperator
${\mathbb D}_o[\rho]$ can be generally written as~\cite{Lindblad-76,
  GKS-76}
\begin{equation}
  {\mathbb D}_o[\rho] = \hat L_o \rho \hat L_o^\dagger - \tfrac{1}{2}
  \big( \rho\, \hat L_o^\dagger \hat L_o + \hat L_o^\dagger \hat L_o
  \rho \big)\,,
  \label{dL}
\end{equation}
where $\hat L_o$ is the Lindblad jump operator associated to the
system-bath coupling scheme.  In the following we will restrict to
homogeneous dissipation mechanisms, preserving translational
invariance, as depicted, for example, in Fig.~\ref{fig:sketch}.  In
quantum optical implementations, the conditions leading to
Eqs.~\eqref{lindblaseq}-\eqref{dL} are typically satisfied~\cite{SBD-16},
therefore this formalism constitutes the standard choice for theoretical
investigations of such kind of systems.

%%%%%%%%%%%%%%%%%%%%%%%%%%%%%%%%%%%%%%%%%%%%%%%%%%%%%%%%%%%%%%%%%%%%%%%
%%%%%%%%%%%%%%%%%%%%%%%%%%%%%%%%%%%%%%%%%%%%%%%%%%%%%%%%%%%%%%%%%%%%%%%

The dissipator ${\mathbb D} \equiv \sum_o {\mathbb D}_o$ typically
drives the system to a steady state, which is generally noncritical,
even when the Hamiltonian parameters are critical.  However, one may
identify a dynamic regime where the dissipation is sufficiently small
to compete with the coherent evolution driven by the critical
Hamiltonian, leading to potentially novel dynamic behaviors.  This is
the target of the present article.  As discussed below, such
low-dissipation regime naturally emerges within a dynamic FSS
framework, assuming many-body systems of linear size $L$ (i.e., of
dimension $L^d$), where the effects of coherent and dissipative
driving terms are somehow {\em measured} in terms of appropriate
powers of $L$.

The paper is organized as follows.
In Sec.~\ref{sec:Framework} we introduce a dynamic FSS framework
addressing the interplay between critical coherent dynamics and dissipation,
for systems described by Lindblad master equations, in which
the coupling with the bath is homogeneous.
Our predictions are verified in Sec.~\ref{sec:Numerics}
for the Kitaev quantum wire subjected to local incoherent particle
losses, pumping, or dephasing.
Finally, in Sec.~\ref{sec:Conclusions} we draw our conclusions.
The Appendix provides technical details on the procedure used
to compute the time trajectories for our model, starting from
the corresponding Lindblad master equation.

\section{Theoretical framework}
\label{sec:Framework}

Our dynamic FSS framework extends the FSS
theory at quantum transitions, already developed at
equilibrium~\cite{SGCS-97, CPV-14, CNPV-14} and in out-of-equilibrium
conditions~\cite{PRV-18, NRV-18} for closed systems.  We assume
that the system Hamiltonian has one relevant parameter $\mu$, whose
tuning toward the point $\mu_c$ develops a quantum critical behavior.
The critical power laws are generally characterized by the
renormalization-group (RG) dimension $y_\mu$ of the relevant parameter
$\bar{\mu}\equiv \mu - \mu_c$ and the dynamic exponent $z$, so that
the diverging length scale behaves as $\xi\approx |\bar{\mu}|^{-\nu}$
with $\nu=1/y_\mu$ and the suppression of the gap (difference of the
two lowest energy levels) as $\Delta \approx \xi^{-z}$. The finite
system size $L$ provides a further relevant length scale.  FSS
is defined by taking the large-$L$ limit, keeping appropriate scaling
variables fixed, such as $\xi/L$~\cite{CPV-14} and $\Delta L^z$ (thus
$\Delta\sim L^{-z}$).
To describe out-of-equilibrium dynamic protocols, for
example arising from a quench of the Hamiltonian control parameter
$\mu$, a further time scaling variable $\theta\propto t \,\Delta$ has
to be introduced~\cite{PRV-18}.  For example, let us consider a sudden
quench of $\mu$ at $t=0$, from $\bar{\mu}_i$ to $\bar{\mu}_f$,
starting from the ground state associated with the initial value
$\bar{\mu}_i$.  We expect that the coherent evolution of a generic
observable, such as the fixed-time correlation $G_{12}$ of two local
operators $\hat{O}_1$ and $\hat{O}_2$ at a distance $x$, undergoes the
asymptotic FSS behavior~\cite{PRV-18}
\begin{eqnarray}
&&G_{12}(x,t,\bar{\mu}_i,\bar{\mu}_f,L) \approx L^{-\varphi} 
{\cal G}(X,\theta,\kappa_i,\kappa_f)\,,
\quad \label{dFSS}\\
&&X \equiv x/L\,,\quad
\theta\equiv t L^{-z}\,,\quad
\kappa_{i/f}\equiv \bar{\mu}_{i/f}  L^{y_\mu}\,,
\label{scavar}
\end{eqnarray}
where $\varphi = y_{1} + y_{2}$ and $y_{i}$ are the RG dimensions of
$\hat{O}_i$.

To account for the effects of the dissipators~\eqref{dL}, we need to
extend the above dynamic FSS theory.  We assume that at $t=0$, beside
quenching the Hamiltonian parameter $\mu$, the dissipation is also
turned on, by effectively, and suddenly, switching the corresponding
effective coupling from zero to some finite value $u>0$. We argue that
the effects of a sufficiently low dissipation can be taken into
account by adding a further dependence on a FSS variable associated
with $u$ in the dynamic FSS Ansatz~\eqref{dFSS}, i.e.~$\gamma = u
L^{\zeta}$ where $\zeta$ is a suitable exponent, to ensure the
substantial balance, thus competition, with the critical coherent
driving.  Since dissipation is predicted to give rise to a relevant
perturbation at the quantum transition, we expect $\zeta>0$.  Thus,
the low-dissipation regime, where the critical coherent dynamics and
dissipation compete, should be characterized by $u\sim
L^{-\zeta}$.
An analogous FSS behavior was put forward to describe
the approach to thermalization of some specific open systems 
close to a quantum transition~\cite{YMZ-14}.

We now argue that the exponent $\zeta$ generally coincides with the
dynamic exponent $z$.  Indeed, we note that the parameter $u$ of the
dissipator in Eq.~\eqref{lindblaseq} plays the role of decay rate,
i.e., of an inverse relaxation time, of the associated dissipative
process~\cite{BP-book}.  Since any relevant time scale $t_s$ at
a quantum transition behaves as $t_s\sim \Delta^{-1}$~\cite{PRV-18},
our working hypothesis is that the dissipation scaling variable
does not involve an independent exponent, but
\begin{equation}
\gamma \equiv u \, L^z\,.
\label{gammaupscaling}
\end{equation}
This implies that, to observe competition between critical
coherent dynamics and dissipation, $u\sim L^{-z}$ must be comparable
with the gap $\Delta\sim L^{-z}$ of
the critical Hamiltonian.  Under the combined effect of coherent and
dissipative driving, the dynamic FSS Ansatz reads
\begin{equation}
  G_{12}(x,t,\bar{\mu}_i,\bar{\mu}_f,u,L) \approx 
  L^{-\varphi} {\cal G}(X,\theta,\kappa_i,\kappa_f,\gamma)\,,
  \label{dFSSd}
\end{equation}
which should be approached in the large-$L$ limit, keeping
the scaling variables $X$, $\theta$,
$\kappa_{i}$, $\kappa_{f}$, and $\gamma$ fixed.  
The convergence to the asymptotic dynamic scaling is
generally characterized by power-law suppressed corrections, 
as usually at continuous quantum transitions.

We conjecture that the Ansatz (\ref{dFSSd}) describes the
low-dissipation regime of quenching protocols for many-body systems at
quantum transitions.  One may also consider an
initial condition given by a Gibbs distribution at temperature $T$.
The dependence on $T$ can be taken into account 
as at equilibrium~\cite{Sachdev-book}, adding a further dependence
on $\tau\equiv T L^z$ in the function ${\cal G}$ of Eq.~(\ref{dFSSd}).
Note that, similarly to the scaling observed at quantum transitions
of closed systems, the dynamic
scaling (\ref{dFSSd}) is expected to be largely independent of the
microscopic properties of the system, that is, it should only depend on
the universality class of the transition and the general properties of
the dissipative mechanism.

We may derive an analogous scaling Ansatz in the infinite-volume limit
$L\to\infty$, keeping the length scale of correlations finite.
In particular, assuming $\bar{\mu}_{i}$ and $\bar{\mu}_{f}$ within the
disordered phase side (thus the quench protocol does not cross the critical
point $\bar{\mu}=0$), for which the ground-state length scales
$\xi_{i,f}$ are large but finite, behaving as
$\xi_{i,f}\sim|\bar{\mu}_{i,f}|^{-\nu}$,
the {\em thermodynamic} $L/\xi_{i,f}\to\infty$ limit of the dynamic FSS
Ansatz~\eqref{dFSSd} can be written as
\begin{equation}
  G_{12} \approx 
  \xi_i^{-\varphi} \widetilde{\cal G}(x/\xi_i,\xi_f/\xi_i,t\,\xi_i^{-z}, u \,\xi_i^{z}).
\end{equation}
A more thorough analysis of the $L\to\infty$ limit,
supported by numerical checks, has been reported in Ref.~\cite{RV-19}.

\section{Kitaev quantum wire coupled to local Markovian baths}
\label{sec:Numerics}

We now present numerical evidence of the above conjecture.  To this
purpose, we consider a Kitaev quantum wire defined by the
Hamiltonian~\cite{Kitaev-01}
\begin{equation}
  \hat H_{\rm K} = - J \sum_{j=1}^L \big( \hat c_j^\dagger \hat
  c_{j+1} + \delta \, \hat c_j^\dagger \hat c_{j+1}^\dagger+{\rm h.c.}
  \big) - \mu \sum_{j=1}^L \hat n_j \,,
  \label{kitaev2}
\end{equation}
where $\hat c_j$ is the fermionic annihilation operator on the $j$th
site of the chain, $\hat n_j\equiv \hat c_j^\dagger \hat c_j$ is the
density operator, and $\delta>0$.  We set $\hslash =1$, and $J=1$ as
the energy scale.  We consider antiperiodic boundary conditions, $\hat
c_{L+1} = - \hat c_1$, and even $L$ for computational convenience.
However, the dynamic scaling scenario applies to general boundary
conditions as well.

The Hamiltonian~\eqref{kitaev2} can be mapped into a spin-1/2 XY
chain, through a Jordan-Wigner transformation~\cite{Sachdev-book}.
It undergoes a continuous quantum transition at $\mu=\mu_c = -2$,
independently of $\delta$, between a disordered ($\mu<\mu_c$) and an
ordered quantum phase ($|\mu|<|\mu_c|$).  This transition belongs to
the two-dimensional Ising universality class~\cite{Sachdev-book},
characterized by the length-scale critical exponent $\nu=1$, related
to the RG dimension $y_\mu = 1/\nu=1$ of the Hamiltonian parameter
$\mu$ (more precisely of the difference $\bar{\mu} \equiv \mu-\mu_c$).
The dynamic exponent associated with the unitary quantum dynamics is
$z=1$.  In the following we set $\delta=1$ (without loss of
generality), such that the corresponding spin model is the quantum
Ising chain
\begin{equation}
  \hat H_{\rm Is} = -\sum_j \hat \sigma^{(3)}_j \hat
  \sigma^{(3)}_{j+1} - g\, \sum_j \hat \sigma^{(1)}_j,
\end{equation}
where $\hat \sigma^{(k)}_j$ are the Pauli matrices and
$g=-\mu/2$~\cite{footnoteIsmo}.

We focus on the dynamic behavior of the Fermi lattice
gas~\eqref{kitaev2} close to its quantum transition, in the presence
of homogeneous dissipation mechanisms following the Lindblad
equation~\eqref{lindblaseq}.  The dissipator
\begin{equation}
  {\mathbb D}[\rho] = \sum_j {\mathbb D}_j[\rho]
\end{equation}
is a sum of local (single-site) terms,
where ${\mathbb D}_j[\rho]$ has the same form as in Eq.~\eqref{dL}
(the index $o$ here corresponds to a lattice site, denoted by $j$).
The onsite Lindblad operators $\hat L_j$ describe the coupling of each
site with an independent bath ${\mathcal B}$, Fig.~\ref{fig:sketch},
and are associated with particle loss (l), pumping (p) and dephasing
(d), respectively~\cite{HC-13, KMSFR-17}:
\begin{equation}
  \hat L_{{\rm l},j} = \hat c_j \,, \quad
  \hat L_{{\rm p},j} = \hat c_j^\dagger \,,\quad
  \hat L_{{\rm d},j} = \hat n_j \,.
  \label{loppe}
\end{equation}
With this choice of dissipators, the full open-system many-body
fermionic master equation enjoys a particularly simple treatment,
enabling a direct solvability for systems with up to thousands of
sites~\cite{Prosen-08, Eisler-11, KMSFR-17}.
Indeed, the dynamics can be written in terms of coupled linear
differential equations, whose number scales linearly with $L$.  We
employ a fourth-order Runge-Kutta method to numerically integrate
them.  Details on the computation of the time trajectories
from the Lindblad Eq.~\eqref{lindblaseq} are reported in the Appendix.
The uniqueness of the eventual steady state has been proven for
the above decay and pumping
operators~\cite{Davies-70, Evans-77, SW-10, Nigro-19}.

%%%%%%%%%%%%%%%%%%%%%%%%%%%%%%%%%%%%%%%%%%%%%%%%%%%%%%%%%%%%%%%%%%%%%%%
\begin{figure}[!t]
  \includegraphics[width=0.98\columnwidth]{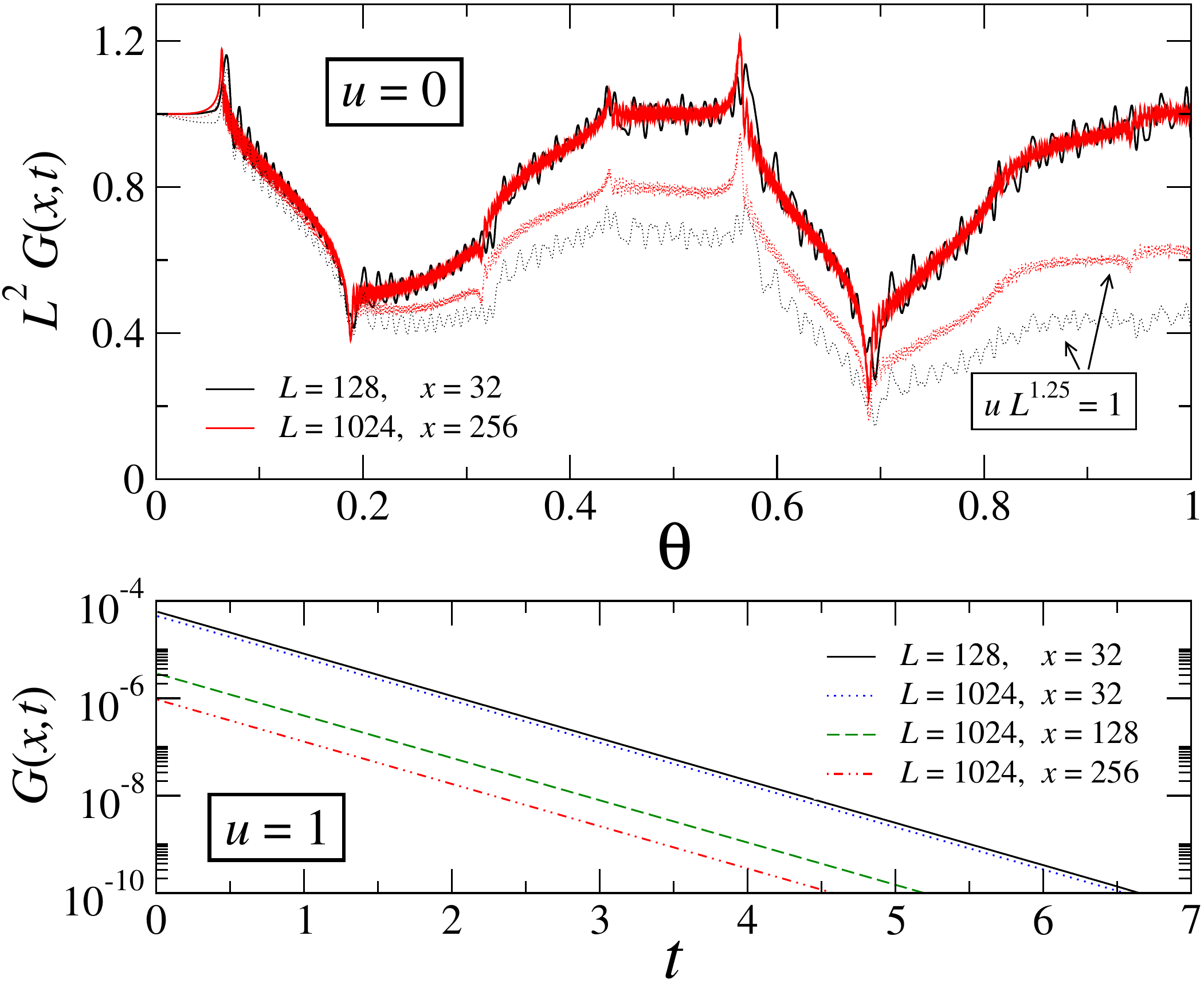}
  \caption{Upper panel: connected density-density correlation function
    $G(x,t)$ for $X= x/L=1/4$, as a function of the rescaled time
    $\theta = t/L$, after a quench from the ground state at the
    critical point $\bar{\mu}_i=0$ to $\bar{\mu}_f$ such that
    $\kappa_f = \bar{\mu}_f L= 2$.  Continuous curves are for a purely
    unitary dynamics ($u=0$), where the dynamic FSS~\eqref{dFSS} is
    verified; dotted curves are in the presence of incoherent particle
    losses, with the dissipative coupling $u$ such that $u L^{1.25} =
    1$.  Lower panel: temporal decay of $G(x,t)$ for a dissipative
    dynamics with $u=1$, at criticality ($\bar{\mu}_i = \bar{\mu}_f =
    0$).  A much faster decay to an uncorrelated state emerges, with a
    slope that asymptotically depends only on $x$.  All curves are
    sufficiently accurate to be considered as practically exact, on
    the scale of this and of all next figures.
    Here and in the next figures, times are
    in units of $\hbar / J$.}
  \label{Check_DissScal2}
\end{figure}
%%%%%%%%%%%%%%%%%%%%%%%%%%%%%%%%%%%%%%%%%%%%%%%%%%%%%%%%%%%%%%%%%%%%%%%

%%%%%%%%%%%%%%%%%%%%%%%%%%%%%%%%%%%%%%%%%%%%%%%%%%%%%%%%%%%%%%%%%%%%%%%
\begin{figure}[!t]
  \includegraphics[width=0.95\columnwidth]{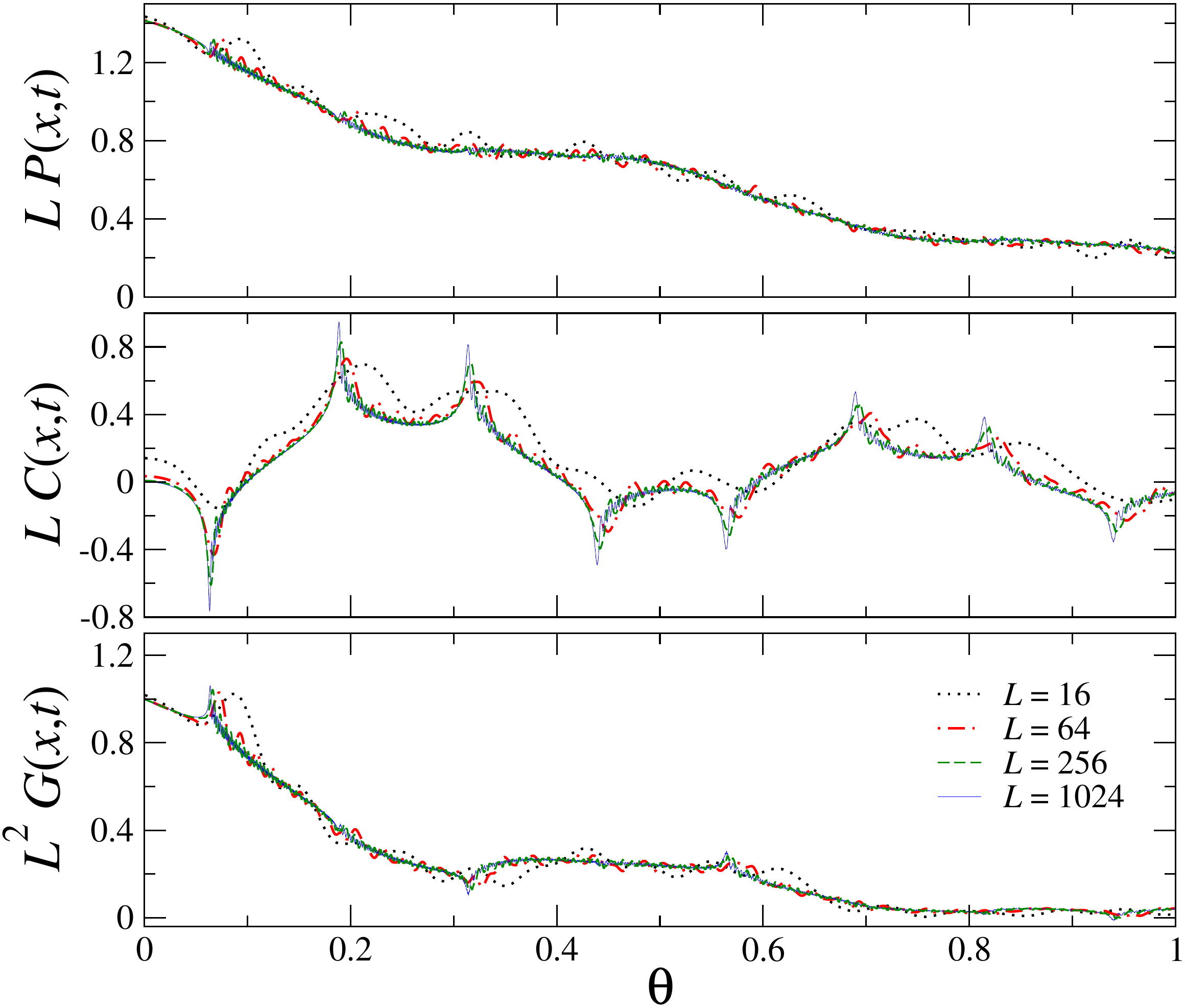}
  \caption{The correlation functions $P(x,t)$ (upper panel), $C(x,t)$
    (central panel), and $G(x,t)$ (lower panel), for $X=1/4$, versus
    $\theta=t/L$.  The system has been driven out of equilibrium
    through a quench from the critical point $\kappa_i =0$ to
    $\kappa_f = 2$, and by the dissipation induced by incoherent
    particle losses, for $\gamma=u L = 1$.  The curves clearly approach a
    scaling function with increasing $L$, thus supporting the dynamic
    FSS in Eq.~\eqref{dFSSd} (here and in Fig.~\ref{Suscept_QD},
    data for $L=256$ are hardly distinguishable from those for $L=1024$).
    Results for other values of $X$, $\kappa_{i,f}$ and $\gamma$ confirm it.}
  \label{Correl_L4_QD}
\end{figure}
%%%%%%%%%%%%%%%%%%%%%%%%%%%%%%%%%%%%%%%%%%%%%%%%%%%%%%%%%%%%%%%%%%%%%%%

Our protocol starts from the ground state of $\hat H_K$ for a generic
$\bar{\mu}_i$, sufficiently small to stay within the critical regime.
%To address the competition between coherent and dissipative dynamics,
We then study the time evolution after a quench of the Hamiltonian parameter
to $\bar{\mu}_f$, and a simultaneous sudden turning on of the dissipation
coupling $u$ (see Appendix for details).  We consider the fixed-time
correlations
\begin{subequations}
\begin{eqnarray}
  P(x,t) & \! = \! & {\rm Tr}[\rho(t)\,(\hat c_j^\dagger \hat c_{j+x}^\dagger +
    \hat c_{j+x} \hat c_{j})],\label{ptf}\\ 
  C(x,t) & \! = \! & {\rm Tr}[\rho(t)\, (\hat c_j^\dagger \hat c_{j+x} + \hat
    c_{j+x}^\dagger \hat c_{j})],\label{gtf}\\ 
  G(x,t) & \! = \! & {\rm Tr}[\rho(t)\, \hat n_j \hat n_{j+x}] - {\rm
    Tr}[\rho(t)\, \hat n_j] \, {\rm Tr}[\rho(t)\, \hat n_{j+x}],
\qquad\label{gntf}
\end{eqnarray}
\end{subequations}
where $j,x \in [1,L/2]$ and $\rho(t)$ is the system's density matrix.
The dynamic FSS behavior of the
observables~\eqref{ptf}-\eqref{gntf} is expected to be given by
Eq.~\eqref{dFSSd}, with $y_\mu=1$, $z=1$. Moreover, $\varphi=1$ for the
correlations $P$ and $C$ (since the RG dimension of the fermionic
operator is $y_{\hat c}=y_{\hat c^\dagger} = 1/2$), while $\varphi=2$ for $G$ (since
$y_{\hat n}=1$).  This scaling scenario should hold for all the considered
dissipation mechanisms, cf. Eq.~\eqref{loppe}. Of course, the
corresponding scaling functions are expected to differ.

Before analyzing the full model, let us
neglect the bath coupling and only consider the unitary dynamics
($u=0$). As is visible from the upper panel of
Fig.~\ref{Check_DissScal2}, for density-density correlations
$G(x,t)$ (continuous curves), the scaling behavior~\eqref{dFSS}
emerges in the large-$L$ limit after the proper rescaling of the pre-
and post-quench control parameter and of time as in
Eq.~\eqref{scavar}. An analogous scenario emerges when $u\ll L^{-z}$;
for example for $u\sim L^{-\zeta}$ with $\zeta>z$, the system
asymptotically converges to the dynamic FSS scenario with $u=0$
(dotted curves in the upper panel), and thus the coherent dynamics
prevails.  Conversely, if the coupling $u$ is switched on and kept
fixed with $L$, the dissipative dynamics overcomes the critical
coherence. Indeed the system exponentially collapses to an
uncorrelated state, in a much shorter time scale (bottom panel ---the
time $t$ has not been rescaled here).  The decay rate only depends on
the distance $x$, thus no scaling behavior emerges.  Similar scenarios
appear whenever $u\gg L^{-z}$.

\begin{figure}[!t]
  \includegraphics[width=0.98\columnwidth]{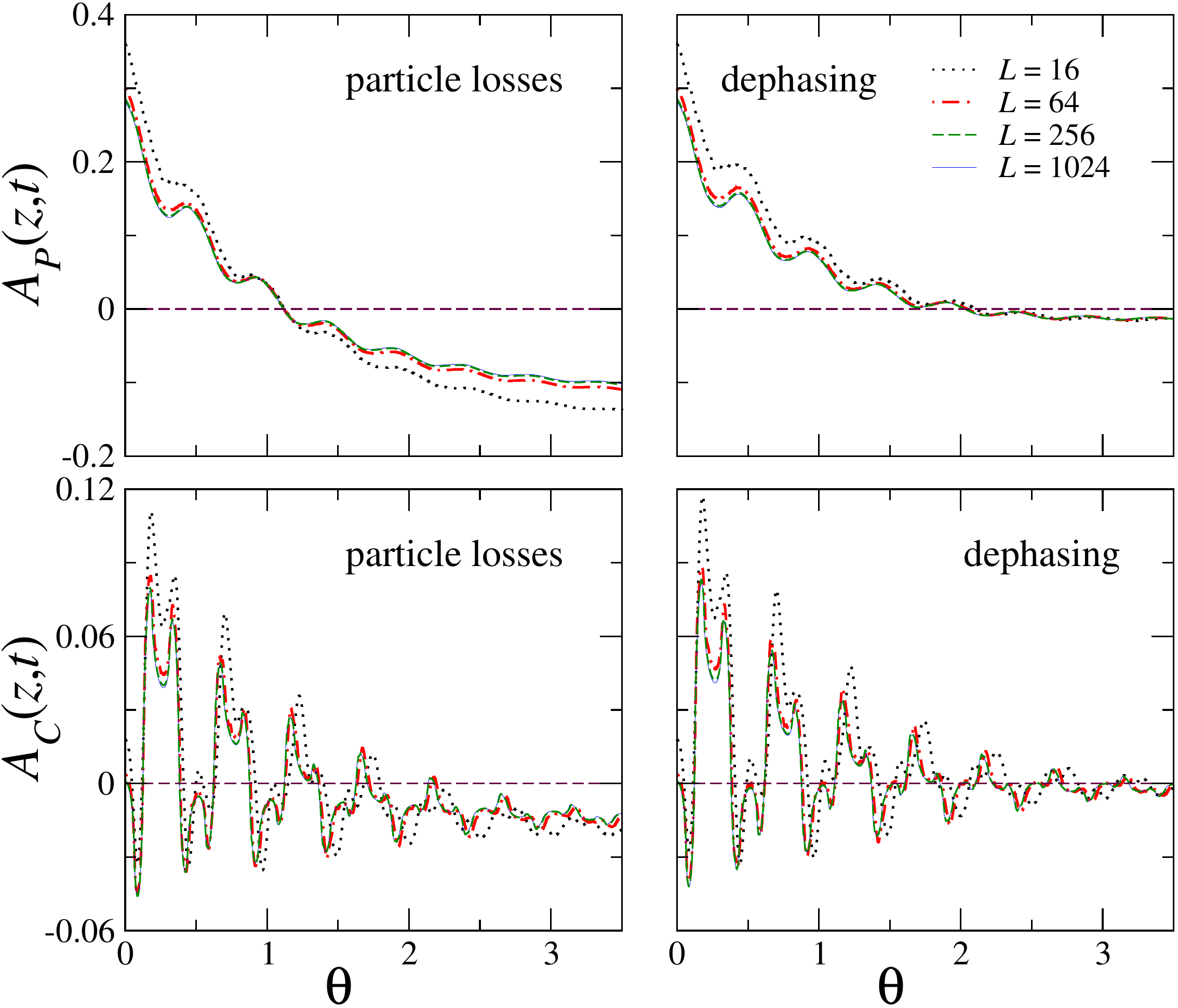}
  \caption{The integrated correlations $A_P$ and $A_C$, for $Z=
    z/L=1/4$, versus $\theta=t/L$.  The system is quenched from
    $\kappa_i = 0$ to $\kappa_f=2$, and dissipation is induced either
    by the presence of incoherent particle losses ($\hat L_{{\rm
        l},j}$, left panels) or by an incoherent dephasing mechanism
    ($\hat L_{{\rm d},j}$, right panels), with $\gamma=1$.  The
    results nicely support Eq.~\eqref{scalaa}, since curves appear to
    converge to a scaling function with increasing $L$.}
  \label{Suscept_QD}
\end{figure}

A nontrivial competition between critical coherence and dissipation
can be only observed for $u\sim L^{-z}$,
cf.~Eq.~\eqref{gammaupscaling}, as shown in Fig.~\ref{Correl_L4_QD}
for a quench protocol in the presence of incoherent particle losses
with rescaled strength $\gamma=u L^{z}=1$.  The dynamic FSS
prediction~\eqref{dFSSd} is clearly verified.  A global check is also
provided by the results shown in Fig.~\ref{Suscept_QD}, for the
integrated correlations
\begin{equation}
  A_P(z,t) = \sum_{x=z}^{L/2}  P(x,t), \quad
  A_C(z,t) = \sum_{x=z}^{L/2} C(x,t), 
  \label{acpdef}
\end{equation}
with $0<z<L/2$, which are expected to scale as
\begin{equation}
  A_{O}(z,t,\bar{\mu}_i, \bar{\mu}_f, u,L) \approx
  L^{1-2y_O} {\cal A}_{O}(Z,\theta,\kappa_i,\kappa_f,\gamma)
  \label{scalaa}
\end{equation}
for $Z \equiv z/L>0$.
Note that this definition cannot be extended to $z\to 0$
(more precisely $Z\to 0$), because the integral of the two-point
function is singular, in that in the critical continuum limit at
equilibrium $C(x)\sim P(x) \sim 1/x$ at small
distance~\cite{Sachdev-book}.

Analogous outcomes are
obtained for the dissipators related to pumping (not shown) and
dephasing (see right panels of Fig.~\ref{Suscept_QD}).  We have
validated our picture also for a lattice gas of free nonrelativistic
fermions, i.e. $\delta =0$ in Eq.~\eqref{kitaev2}, which undergoes a
quantum transition lying in a different universality class, with
dynamic exponent $z=2$ (not shown).

%%%%%%%%%%%%%%%%%%%%%%%%%%%%%%%%%%%%%%%%%%%%%%%%%%%%%%%%%%%%%%%%%%%%%%%
%%%%%%%%%%%%%%%%%%%%%%%%%%%%%%%%%%%%%%%%%%%%%%%%%%%%%%%%%%%%%%%%%%%%%%%

\section{Conclusions}
\label{sec:Conclusions}

In summary, our findings confirm the existence of a dynamic regime
characterized by the competition between critical coherent and
dissipative dynamics, supporting the scaling behaviors put forward
within the dynamic FSS framework.  
We will report elsewhere a more
thorough discussion of the numerical results, their convergence rate
[which is generally $O(L^{-1})$ in the critical Kitaev model], the particular
features of the scaling curves, such as the emerging
spikes in the rescaled time $\theta$ (reminiscent of the behavior at
dynamical phase transitions~\cite{Heyl-18}, see
Figs.~\ref{Check_DissScal2} and~\ref{Correl_L4_QD}), and the
asymptotic behaviors in the
large-$\theta$ limit.
The dynamic FSS framework can be also used to study other protocols in
the presence of dissipation, for example when slowly changing the
Hamiltonian parameters across a quantum transition.

The arguments leading to the above scaling scenario are quite
general. Analogous phenomena are expected to develop in any
homogeneous $d$-dimensional many-body system at a continuous quantum
transition, whose Markovian interaction with the bath can be
described by local or extended dissipators within a Lindblad
equation~\eqref{lindblaseq}.  The regime showing competition of
critical coherent and dissipative dynamics is realized when the
dissipation parameter $u$ scales as the gap $\Delta$ of the
Hamiltonian of the many-body system, i.e.,
\begin{equation}
  u\sim \Delta\,.
\end{equation}
Since at a quantum transition $\Delta\sim L^{-z}$, this is a
low-dissipation regime. This reflects the fact that at a quantum
transition the perturbation arising from dissipation is always
relevant, such the temperature at equilibrium~\cite{Sachdev-book,
  SGCS-97, CPV-14}.  Therefore, when $u\gg \Delta$, critical coherent
fluctuations do not survive dissipation. These arguments should also
apply to non-Markovian system-bath couplings~\cite{DA-17} (not
described by Lindblad equations), replacing $u$ with the parameter
controlling the decay rate.

This dynamic scenario has been checked within fermion wires, cf.
Eq.~\eqref{kitaev2}, in the presence of local dissipation mechanisms
associated with the Lindblad operators~\eqref{loppe}.  Further studies
would serve to achieve a conclusive validation of our competition
theory, for other many-body systems and/or dissipation mechanisms,
including nonlocal ones~\cite{VWC-09, DRBZ-11, LPK-16}.  Further
interesting issues may concern quantum thermodynamic
properties~\cite{Thermo-book, Thermo-book2, CF-16} in the competition regime.

Other issues worth being investigated concern the emergence, and
characterization, of analogous competition scaling phenomena at
first-order quantum transitions, for which dynamic FSS frameworks
have been also developed~\cite{PRV-18}, and new features may arise,
like a particular sensitivity on the type of
boundary conditions~\cite{CNPV-14,PRV-18c}.

We finally mention that some experimental breakthroughs have been
recently achieved in the control of dissipative quantum many-body
dynamics, through different platforms, such as Rydberg atoms or
circuit-QED technology. For example, a quantum critical behavior in
such out-of-equilibrium context was reported~\cite{CRWAW-13,
  TNDTT-17, FSLKH-17}.  These studies encourage the verification of
our competition theory, using a limited (relatively small, say, few
tens) amount of controlled objects, which may already suffice to
highlight some signatures of dynamic scaling.

\appendix

\section{Solution of Eq.~\eqref{lindblaseq} for our Fermi lattice gas model}

It is useful to first distinguish between the different
schemes of system-bath coupling employed in this work.
Specifically, if the dissipation is linear in the creation and/or annihilation operators,
as is the case for incoherent particle losses ($\hat{L}_{{\rm l},\,j}=\hat{c}_j$)
or pumping ($\hat{L}_{{\rm p},\,j} = \hat{c}^\dagger_j$),
the corresponding driven-dissipative quantum dynamics can be exactly solved
using an analogous strategy as for standard quadratic Fermi models,
which reduces the exponential complexity of the problem to a polynomial one.
In contrast, a different method has to be adopted
for a dephasing mechanism ($\hat{L}_{{\rm d},\,j}=\hat{n}_j$), 
where, although the full dynamics cannot be simply obtained, it is however possible
to track the time evolution of certain expectation values, using a polynomial
amount of resources.

In this respect, this appendix contains an excerpt of some technicalities
which have been already detailed in Refs.~\cite{Prosen-08, HC-13, KMSFR-17}.
These are reported here for the sake of clarity, and in order to make our discussion
self consistent and useful to anyone who needs to reproduce our results.
On top of that, we also provide additional details on the specific observables discussed
in Sec.~\ref{sec:Numerics}, and on the implementation of the antiperiodic
boundary conditions for our model.

\subsection{Quantum dynamics in the presence of incoherent losses or pumping}
\label{app:1}

For $\hat{L}_{j}=\hat{c}^{(\dagger)}_j$, the dissipator
${\mathbb D}[\rho] = \sum_o {\mathbb D}_o[\rho]$ in Eq.~\eqref{lindblaseq} turns out
to be quadratic in the fermionic creation and annihilation operators
(notice that the index $o$ here corresponds to a lattice site, denoted by $j$).
The same feature holds for the Kitaev Hamiltonian in Eq.~\eqref{kitaev2}.
As a consequence and since the system is translationally invariant, it is useful to
perform a Fourier transformation applied to creation/annihilation operators
for fermions on the chain~\cite{Sachdev-book}:
\begin{equation}
  \label{eq:FourierTranformAnnihilation}
  \hat{c}_j = \frac{e^{-i\pi/4}}{\sqrt{L}}\sum_k \hat{c}_{k}e^{i k j}, \qquad
  (j = 1,\ldots, L).
\end{equation}
This transformation preserves the fermionic anticommutation rules, that is,
\begin{subequations}
  \begin{align}
    \{ \hat c_j, \hat c_l \} \! = \! 0 , \;\; & \{ \hat c_j^\dagger, \hat c_l \} \! = \! \delta_{j,l}, &\mbox{real space,} \qquad\\
    \{ \hat c_k, \hat c_q \} \! = \! 0 , \;\; & \{ \hat c_k^\dagger, \hat c_q \} \! = \! \delta_{k,q}, &\mbox{momentum space.}
  \end{align}
  \label{eq:anticomm}
\end{subequations}
Considering, without loss of generality, an even number $L$ of sites in the chain,
antiperiodic boundary conditions can be enforced by choosing the following set of momenta:
\begin{equation}
  \label{eq:oddLmomenta}
  k = \Big\{ \pm \frac{\pi}{L}(2n+1) \Big\} , \quad n=0,1,\cdots, L/2-1 .
\end{equation}
Indeed, adopting such choice and using the definition
in Eq.~\eqref{eq:FourierTranformAnnihilation}, it is easy to see that 
\begin{equation}
  \hat c^{(\dagger)}_{j+L} = - \hat c^{(\dagger)}_j, \qquad \forall j \in [1,L].
\end{equation}

The density operator $\rho(t)$ at time $t=0$ is taken as the ground state
of $\hat H_{\rm K}$. This can be cast in a tensor product form, after going
in momentum space:
\begin{equation}\label{eq:rhofactorisedzero}
  \rho(0) = \bigotimes_{k>0} \rho_{k}(0).
\end{equation}
Here $\rho_k(0)$ denotes the restricted density operator describing the configuration
of the $k$-sector (with $k>0$), that is, the sector containing contributions associated
to excitations having momentum $\pm \vert k\vert$.
Due to the structure of the Lindblad master equation, $\rho(t)$ is factorized in momentum space
for any time $t$. As a consequence, the behavior in each $k$-sector
can be determined by solving a differential system having the following structure
(in units of $\hslash = 1$):
\begin{equation}
  \frac{d}{dt}\rho_k(t)=-i \big[ \hat{H}_k,\rho_k(t) \big] + u \, \mathbb{D}[\rho_k], \quad k>0.
  \label{eq:kspace}
\end{equation}
The Hamiltonian
\begin{equation}\label{eq:quadratichamiltoniankspace}
\hat{H}_k=\left(
\begin{matrix}
  0 		 &      0      &     0 		& 2 \,|\!\sin k|  \\
  0 		 & -2 f_k(\mu) &     0 		&   0             \\
  0 	 	 &      0      &   -2 f_k(\mu) 	&   0		  \\
  2 \,|\!\sin k| &      0      &     0		&  -4f_k(\mu)
\end{matrix}\right),
\end{equation}
with $f_k(\mu) = \mu/2 + \cos k$ [we put $J=\delta=1$ in Eq.~\eqref{kitaev2}],
governs the dynamics in the four dimensional state basis
$\{\vert 0_{k}\rangle,\,\vert 1 _{k}\rangle,\,\vert 1_{-k} \rangle,\,\vert 1 _{k},\,1_{-k} \rangle \}$.
The dissipator in the corresponding $k$-sector violates the fermion parity;
for the case of homogeneous particle losses (i.e., $\hat L_j = \hat c_j$)
this is given by
\begin{eqnarray}
  \mathbb{D}[\rho_{k}] & = & \hat{c}_{k}\rho_{k}\hat{c}^{\dagger}_{k}-\tfrac{1}{2}(\hat{n}_{k}\rho_{k}+\rho_{k}\hat{n}_{k})+ \nonumber \\
  & + & \hat{c}_{-k}\rho_{k}\hat{c}^{\dagger}_{-k}-\tfrac{1}{2}(\hat{n}_{-k}\rho_{k}+\rho_{k}\hat{n}_{-k}).
  \label{eq:Dk}
\end{eqnarray}
A very similar expression for $\mathbb{D}[\rho_k]$ holds, with analogous properties
as those for Eq.~\eqref{eq:Dk}, in the case of homogeneous particle pumping
(i.e., $\hat L_j = \hat c^\dagger_j$), provided these substitutions are applied in the above equation:
\begin{subequations}
  \begin{eqnarray}
    \hat c_k \to \hat c^\dagger_k, &\quad &\hat c_{-k} \to \hat c^\dagger_{-k}, \\
    \hat c^\dagger_k \to \hat c_k, &\quad &\hat c^\dagger_{-k} \to \hat c_{-k}, \\
    \hat n_k \to \hat c_k \hat c^\dagger_k, &\quad &\hat n_{-k} \to \hat c_{-k} \hat c^\dagger_{-k} .
  \end{eqnarray}
\end{subequations}

Once the structure of all the $\rho_{k}(t)$ matrices is determined by explicitly
solving Eq.~\eqref{eq:kspace} in the corresponding four-dimensional Hilbert
$k$-subspace (recall that $k>0$), the time evolution
of any observable can be computed simply by inverting the mapping
in Eq.~\ref{eq:FourierTranformAnnihilation}.
Indeed, given an observable $\hat{\mathcal{O}}(\{\hat{c}_j\},\{\hat{c}_j^{\dagger}\})$
in real space, its explicit time evolution is obtained by moving into momentum space:
$\hat{\mathcal{O}} (\{\hat{c}_{k}\},\{\hat{c}^{\dagger}_{k}\})$ and then considering
the average
\begin{equation}
  \langle \hat{\mathcal{O}} \rangle(t) = \mbox{Tr} \Big[ \hat{\mathcal{O}} \big(
    \{\hat{c}_{k}\},\{\hat{c}^{\dagger}_{k}\} \big) \, \bigotimes_{k>0}\rho_{k}(t) \Big].
\end{equation}
In the present case, we also have that
\begin{equation}\label{eq:zeroamplitudes}
  \langle \hat{c}_{k}\rangle(t) = \langle \hat{c}^{\dagger}_{k}\rangle(t) = 0,
  \quad \forall t,\,k \,.
\end{equation}
This can be easily shown by considering the equations of motion for such amplitudes. As a consequence, the only operators that can have non-zero expectation value are those corresponding to products of an even number of fermionic operators in each $k$-subspace. In all the other cases, the expectation values are zero, due to Eq.~\ref{eq:zeroamplitudes} and to the anticommutation rules~\eqref{eq:anticomm}.

Let us now explicitly consider the pairing correlation function $P(x,t)$
[see Eq.~\eqref{ptf}], that is, 
\begin{equation}
  P(x,t) = \langle \hat{c}^{\dagger }_{j}\hat{c}^{\dagger}_{j+x}\rangle(t)+\langle \hat{c}_{j+x}\hat{c}_{j}\rangle(t) .
  \label{eq:Pcorr}
\end{equation}
Such quantity in momentum space is given by
\begin{equation}
  P(x,t)=\bigg[ \frac{e^{i\pi/2}}{L} \!\!
  \sum_{k,q} e^{-i [ k j + q (j+x) ]}\langle \hat{c}^{\dagger}_{k}\hat{c}^{\dagger}_{q}\rangle(t) \bigg] + {\rm h.c.}
  \label{eq:Pmom}
\end{equation}
Due to the constraint listed above, we also have that 
\begin{equation}
\langle \hat{c}_{k}\hat{c}_{q}\rangle(t) \neq 0 \quad \Longleftrightarrow \quad k=-q .
\end{equation}
As a consequence, the expression in Eq.~\eqref{eq:Pmom} further simplifies into 
\begin{equation}
  P(x,t) = \bigg[ \frac{e^{i\pi/2}}{L}\sum_{k} e^{i k x} \langle
    \hat{c}^{\dagger}_{k}\hat{c}^{\dagger}_{-k}\rangle(t) \bigg] + {\rm h.c.},
\end{equation}
which can be eventually written in a more compact form as
\begin{equation}
  P(x,t) = -\frac{2}{L}\sum_{k > 0}\sin(k x) \left[
    \langle c^{\dagger}_{k}c^{\dagger}_{-k}\rangle(t) + {\rm h.c.} \right] .
\end{equation}

By exploiting the same strategy, it is possible to decompose any mean value
as a combination of amplitudes that involve expectation values in momentum space.
For instance, if one considers the correlation function $C(x,t)$ of Eq.~\eqref{gtf},
one finds the expression
\begin{equation}
  C(x,t) = \frac{2}{L}\sum_{k > 0}\cos(k x) \left[
    \langle \hat{c}^{\dagger}_{k}\hat{c}_{k}\rangle(t) +
    \langle \hat{c}^{\dagger}_{-k}\hat{c}_{-k}\rangle(t) \right] .
\end{equation}
As the number of operators in real space increases, the structure in momentum space
becomes more cumbersome. This is the case for the four-point connected density-density
operator $G(x,t)$ of Eq.~\eqref{gntf}, which can be expressed as 
\begin{widetext}
\begin{align}
  & G(x,t) = \frac{\delta_{x,0}}{L}\sum_{k>0}\left[\langle \hat{n}_k\rangle + \langle \hat{n}_{-k}\rangle\right]
  +\frac{2}{L^2}\sum_{k>0} \Big\{ \big[ \cos(2 k x) -1 \big]
  \big[ \langle \hat{n}_k \rangle \langle \hat{n}_{-k} \rangle - \langle \hat{n}_k \hat{n}_{-k} \rangle
  + \langle \hat{c}^{\dagger}_k \hat{c}^{\dagger}_{-k}\rangle\langle \hat{c}_{-k} \hat{c}_{k}\rangle \big] \Big\}
  \nonumber \\
  & + 
  \frac{4}{L^2} \bigg\{ \! \sum_{k>0}\sin (k x) \langle \hat{c}^{\dagger}_{k}\hat{c}^{\dagger}_{-k} \rangle \! \bigg\}
  \bigg\{ \! \sum_{k>0}\sin (k x) \langle \hat{c}_{-k}\hat{c}_{k} \rangle \! \bigg\} 
  \! - \! \frac{1}{L^2} \bigg\{ \! \sum_{k>0} \left[ e^{ikx} \langle \hat{n}_k \rangle
    +e^{-ikx} \langle \hat{n}_{-k} \rangle \right] \! \bigg\}
   \bigg\{ \! \sum_{k>0} \left[ e^{-ikx} \langle \hat{n}_k \rangle
    + e^{ikx} \langle \hat{n}_{-k} \rangle \right] \! \bigg\},
\end{align}
\end{widetext}
where, for the ease of compactness, we have omitted the time dependence of all
the expectation values.

We end up by mentioning that antiperiodic boundary conditions are automatically
guaranteed by adopting the choice of momenta $k$ written in Eq.~\eqref{eq:oddLmomenta}.
The expressions we have reported for the correlations in $k$-space
correspond to measuring them in real space within the chain length, that is,
by taking $j, x \in [1, L/2]$ in Eq.~\eqref{eq:Pcorr} (and similar).
Otherwise, a minus sign would appear each time the boundary is crossed an
odd number of times, since
\begin{equation}
  \hat c^{(\dagger)}_{j + mL} = (-1)^{m} \hat c^{(\dagger)}_j, \quad \forall j \in [1,L].
\end{equation}

Notice also that the following symmetries always hold (where $x \in [1,L]$),
due to antiperiodic boundaries:
\begin{subequations}
  \begin{eqnarray}
    P(x,t) & = & P(L-x,t), \\
    C(x,t) & = & -C(L-x,t), \\
    G(x,t) & = & G(L-x,t).
  \end{eqnarray}
\end{subequations}

\subsection{Quantum dynamics in the presence of dephasing}
\label{app:2}

Unfortunately, the quantum dynamics of the fermionic Kitaev chain in the presence
of dephasing Lindblad terms
$\hat L_{{\rm d},j} = \hat n_j = \hat c^\dagger_j \hat c_j$ does not factorize
in momentum space, since the dissipator ${\mathbb D}[\rho]$ now becomes quartic
in the creation/annihilation operators.
As a consequence, the method described in App.~\ref{app:1} cannot be exploited and,
in general, an exact solution in terms of a polynomial scaling with $L$ cannot be obtained.
Nonetheless, one could pay attention only to the time evolution of certain observables
of interest.
We recall that, in order to determine the behavior of a given time-dependent
expectation value $\langle \hat{\mathcal{O}}\rangle(t)$, one needs to solve
the following differential equation~\cite{BP-book}
\begin{equation}\label{eq:heisenbergpicture}
  \frac{d}{dt} \hat{\mathcal{O}}= i \big[ \hat{H},\hat{\mathcal{O}} \big] + u \, \tilde{\mathbb{D}}[\hat{\mathcal{O}}],
\end{equation}
where 
\begin{equation}
  \tilde{\mathbb{D}}[\hat{\mathcal{O}}]=\sum_{j}\left[\hat{L}^{\dagger}_j \hat{\mathcal{O}} \hat{L}_j
    - \tfrac{1}{2} \big\{ \hat{L}^{\dagger}_j \hat{L}_j,\hat{\mathcal{O}} \big\} \right]
\end{equation}
denotes the dissipator in the Heisenberg picture.

Solving Eq.~\eqref{eq:heisenbergpicture} for a many-body system is generally an hard task,
unless explicit constructions as the one reported in App.~\ref{app:1} are possible.
Indeed, the time evolution of a given operator usually depends also on that
of other observables. As a consequence, solving a single equation of motion
actually requires to deal with a number of differential equations that usually
grows exponentially with the system size $L$.

In the present case, for the two-point observables $P(x,t)$ and $C(x,t)$ of
Eqs.~\eqref{ptf} and~\eqref{gtf}, it is however possible to find a closed set
of equations of motion, whose dimension grows only polynomially with increasing $L$.
Such set is given by all the \emph{quadratic} observables in the fermionic operators.
In such case, the time evolution of any two point amplitude
[as is the case for $P(x,t)$ and $C(x,t)$], can be rephrased in terms
of the behavior of the following 4$L$ amplitudes: 
\begin{subequations}
\begin{eqnarray}
  \label{eq:Aamplitude}
  \langle \mathcal{\hat A}_x \rangle(t) & \equiv & \langle \hat{c}_j \hat{c}_{j+x} \rangle(t), \\
  \label{eq:Bamplitude}
  \langle \mathcal{\hat B}_x \rangle(t) & \equiv & \langle \hat{c}_j \hat{c}^{\dagger}_{j+x} \rangle(t), \\
  \label{eq:Camplitude}
  \langle \mathcal{\hat C}_x \rangle(t) & \equiv & \langle \hat{c}^{\dagger}_j \hat{c}_{j+x}\rangle(t), \\
  \label{eq:Damplitude}
  \langle \mathcal{\hat D}_x \rangle(t) & \equiv & \langle \hat{c}^{\dagger}_j \hat{c}^{\dagger}_{j+x}\rangle(t) .
\end{eqnarray}
\label{eq:amplitudes}
\end{subequations}
Here we always suppose that $j \in [1,L]$ and $j+x \in [1,L]$,
such that the boundaries of the chain are never crossed.
By plugging these operators in Eq.~\eqref{eq:heisenbergpicture}, one arrives
at the following set of $4L$ coupled differential equations governing
the time evolution of the corresponding amplitudes in Eqs.~\eqref{eq:amplitudes}:
\begin{widetext}
  \begin{subequations}
    \begin{eqnarray}
      \frac{d\mathcal{\hat A}_x}{dt} & = & 2 i (\mathcal{\hat A}_{x-1}+\mathcal{\hat A}_{x+1})
      + i (\mathcal{\hat C}_{x-1} - \mathcal{\hat C}_{x+1}) - i (\mathcal{\hat B}_{x-1} - \mathcal{\hat B}_{x+1})
      + 2 i \mu \mathcal{\hat A}_x - u \, \mathcal{\hat A}_x, \label{eq:eomA} \\
      \frac{d\mathcal{\hat B}_x}{dt} & = & i (\mathcal{\hat D}_{x-1} - \mathcal{\hat D}_{x+1})
      + i (\mathcal{\hat A}_{x-1} - \mathcal{\hat A}_{x+1}) - u \, \mathcal{\hat B}_x(1-\delta_{x,0}), \label{eq:eomB} \\
      \frac{d \mathcal{\hat C}_x}{dt} & = & -i ( \mathcal{\hat D}_{x-1} - \mathcal{\hat D}_{x+1})
      - i(\mathcal{\hat A}_{x-1} - \mathcal{\hat A}_{x+1}) - u \, \mathcal{\hat C}_x(1-\delta_{x,0}), \label{eq:eomC} \\
      \frac{d\mathcal{\hat D}_x}{dt} & = & -2 i (\mathcal{\hat D}_{x-1} + \mathcal{\hat D}_{x+1})
      + i ( \mathcal{\hat C}_{x-1} - \mathcal{\hat C}_{x+1}) - i (\mathcal{\hat B}_{x-1} - \mathcal{\hat B}_{x+1})
      -2 i \mu \mathcal{\hat D}_x - u \, \mathcal{\hat D}_x . \label{eq:eomD}
    \end{eqnarray}
    \label{eq:diffsys}
  \end{subequations}
\end{widetext}

Since one is looking for the structure of four different kinds of amplitudes and $x$ can take values
from $0$ to $L-1$, the number of coupled equations is $4L$.
Indeed, due to the presence of first-neighbor coupling terms in the Kitaev chain,
the amplitudes corresponding to operators at distance $x$ are related to those at distance $x+1$ and $x-1$.
However, due to translational invariance and fermionic statistics, such amplitudes possess symmetry properties
that enable to reduce the amount of coupled equations of motion to be solved.
Indeed the following relations hold ($x \in [0,L-1]$):
\begin{subequations}
\begin{eqnarray}
  \mathcal{\hat A}_x & = & - \mathcal{\hat A}_{-x} , \\
  \mathcal{\hat B}_x & = & - \mathcal{\hat C}_{-x} + \delta_{x,0} , \\
  \mathcal{\hat D}_x & = & - \mathcal{\hat D}_{-x} .
\end{eqnarray}
\end{subequations}
In addition, by exploiting antiperiodic boundary conditions, we have that ($y \in [0,L/2-1]$):
\begin{align}
  &\mathcal{\hat A}_{L/2+y} = \hat c_{L/2} \hat c_{L+y} =
  - \hat c_{L/2} \hat c_y = \hat c_y \hat c_{L/2} = \mathcal{\hat A}_{L/2-y}, \nonumber \\
  &\mathcal{\hat B}_{L/2+y} = \hat c_{L/2} \hat c^\dagger_{L+y} = 
  - \hat c_{L/2} \hat c^\dagger_y = \hat c^\dagger_y \hat c_{L/2} = \mathcal{\hat C}_{L/2-y}, \nonumber \\
  &\mathcal{\hat C}_{L/2+y} = \hat c^\dagger_{L/2} \hat c_{L+y} = 
  - \hat c^\dagger_{L/2} \hat c_y = \hat c_y \hat c^\dagger_{L/2} = \mathcal{\hat B}_{L/2-y}, \nonumber \\
  &\mathcal{\hat D}_{L/2+y} = \hat c^\dagger_{L/2} \hat c^\dagger_{L+y} = 
  - \hat c^\dagger_{L/2} \hat c^\dagger_y = \hat c^\dagger_y \hat c^\dagger_{L/2} = \mathcal{\hat D}_{L/2-y},
\end{align}
where we have plugged $j=L/2$ in Eqs.~\eqref{eq:amplitudes}.
It is thus clear that the full problem for the above two-point correlators is actually $(2L+2)$-dimensional,
since it is sufficient to write the corresponding coupled equations for the operators: $\mathcal{\hat A}_{x \in [1,L/2]}$,
$\mathcal{\hat B}_{x \in [0,L/2]}$, $\mathcal{\hat C}_{x \in [0,L/2]}$, $\mathcal{\hat D}_{x \in [1,L/2]}$.
Notice also that one trivially has $\mathcal{\hat A}_0 = \mathcal{\hat D}_0 = 0$.

The initial conditions for the differential system~\eqref{eq:diffsys}
correspond to the expectation values of such operators evaluated on the ground state
of the Kitaev chain for a given value of the control parameter $\mu = \mu_i$,
and can be immediately found by means of a Bogoliubov transformation in real space,
which generalizes the standard procedure in $k$-space to nonhomogeneous quadratic systems.
Once the time evolution of the amplitudes~\eqref{eq:amplitudes} is determined, the behavior
of the two-point observables $P(x,t)$ and $C(x,t)$ for $x \in [1,L/2]$ can be easily accessed by noticing that
\begin{subequations}
  \begin{eqnarray}
    P(x,t) & = & \langle \mathcal{\hat D}_x \rangle(t) - \langle \mathcal{\hat A}_x \rangle(t)
    = 2 \, {\rm Re} \big[ \langle \mathcal{\hat D}_x \rangle(t) \big], \qquad \\
    C(x,t) & = & \langle \mathcal{\hat C}_x \rangle(t) - \langle \mathcal{\hat B}_x \rangle(t).
  \end{eqnarray}
\end{subequations}


\begin{thebibliography}{99}
\bibitem{footnoteIsmo} Note however that the non-local Jordan-Wigner
  transformation of the Ising chain with periodic or antiperiodic
  boundary conditions does not map into the fermionic
  model~\eqref{kitaev2} with periodic or antiperiodic boundary
  conditions. Indeed further considerations
  apply~\cite{Katsura-62,Pfeuty-70}, leading to a less straightforward
  correspondence, depending on the parity of the particle number
  eigenvalue.  Therefore, although their bulk behaviors in the
  infinite-volume limit, and thus their phase diagram, are analogous,
  the resulting FSS functions differ, since they depend on the choice
  of the boundary conditions.
\end{thebibliography}
\end{document}